\documentclass[%
 reprint,
superscriptaddress,
 amsmath,amssymb,
 aps,
 prl,
]{revtex4-2}

\usepackage{graphicx}% Include figure files
\usepackage{dcolumn}% Align table columns on decimal point
\usepackage{bm}% bold math
\usepackage{xcolor}
\usepackage{hyperref}
\usepackage{braket}
\usepackage{amsmath, amssymb}
\usepackage{bbold}
\usepackage{soul}
\newcommand\rst{\bgroup\markoverwith{\textcolor{red}{\rule[0.5ex]{2pt}{0.4pt}}}\ULon}

\begin{document}

\preprint{APS/123-QED}

\title{Observation of two-mode squeezing in a traveling wave parametric amplifier}
    \author{Martina Esposito}
    \thanks{These authors contributed equally to the work.}
    \affiliation{Univ. Grenoble Alpes, CNRS, Grenoble INP, Institut N\'eel, 38000 Grenoble, France}
    \affiliation{CNR-SPIN Complesso di Monte S. Angelo, via Cintia, Napoli, 80126, Italy}
    \author{Arpit Ranadive} 
    \thanks{These authors contributed equally to the work.}
    \affiliation{Univ. Grenoble Alpes, CNRS, Grenoble INP, Institut N\'eel, 38000 Grenoble, France}
    \author{Luca Planat}
    \affiliation{Univ. Grenoble Alpes, CNRS, Grenoble INP, Institut N\'eel, 38000 Grenoble, France}
    \author{Sébastien Léger}
    \affiliation{Univ. Grenoble Alpes, CNRS, Grenoble INP, Institut N\'eel, 38000 Grenoble, France}
    \author{Dorian Fraudet}
    \affiliation{Univ. Grenoble Alpes, CNRS, Grenoble INP, Institut N\'eel, 38000 Grenoble, France}
    \author{Vincent Jouanny}
    \affiliation{Univ. Grenoble Alpes, CNRS, Grenoble INP, Institut N\'eel, 38000 Grenoble, France}
    \author{Olivier Buisson}
    \affiliation{Univ. Grenoble Alpes, CNRS, Grenoble INP, Institut N\'eel, 38000 Grenoble, France}
    \author{Wiebke Guichard}
    \affiliation{Univ. Grenoble Alpes, CNRS, Grenoble INP, Institut N\'eel, 38000 Grenoble, France}
    \author{C\'ecile Naud}
    \affiliation{Univ. Grenoble Alpes, CNRS, Grenoble INP, Institut N\'eel, 38000 Grenoble, France}
    \author{José Aumentado}
    \affiliation{National Institute of Standards and Technology, Boulder, Colorado, 80305, USA}
    \author{Florent Lecocq}
    \affiliation{National Institute of Standards and Technology, Boulder, Colorado, 80305, USA}
    \author{Nicolas Roch}
    \affiliation{Univ. Grenoble Alpes, CNRS, Grenoble INP, Institut N\'eel, 38000 Grenoble, France}

% \date{\today}

\begin{abstract}
Traveling wave parametric amplifiers (TWPAs) have recently emerged as essential tools for broadband near quantum-limited amplification. However, their use to generate microwave quantum states still misses an experimental demonstration. In this letter, we report operation of a TWPA as a source of two-mode squeezed microwave radiation. We demonstrate broadband entanglement generation between two modes separated by up to $400$ MHz by measuring logarithmic negativity between 0.27 and 0.51 and collective quadrature squeezing below the vacuum limit between 1.5 and 2.1 dB. This work opens interesting perspectives for the exploration of novel microwave photonics experiments with possible applications in quantum sensing and continuous variable quantum computing. 
\end{abstract}

\maketitle

Frequency conversion and wave mixing processes in nonlinear media allow manipulation and control of the electromagnetic radiation \cite{Boyd} and are extensively used for a broad range of applications including, for example, coherent high harmonics generation \cite{Kim2008}, nonlinear spectroscopy \cite{Geissbuehler2012}, nonlinear imaging \cite{Dacosta2014}  and quantum optics experiments for the generation of entanglement and squeezing \cite{Barz2010,Castellanos-Beltran2008}. %\\
In the last two decades, it became possible to tailor such nonlinear interactions by engineering artificial media with specifically designed nonlinear properties: nonlinear meta-materials \cite{Lapine2014,Jung2014,Krasnok2018}. 
Superconducting quantum circuits based on Josephson junctions recently gained a key role in this framework since they can be used to engineer strong nonlinearities without dissipation.

Josephson junctions based nonlinar meta-materials \cite{Lapine2014,Jung2014} have been successfully implemented as near quantum-noise-limited traveling wave parametric amplifiers (TWPAs) \cite{Macklin2015, White2015, Zorin2018,Miano2019,Planat2020b,Ranadive2021}. However, the potential of these devices goes far beyond amplification: since they offer large bandwidth and flexible customization of the desired nonlinear response,
they have been identified as very promising for the generation of two-mode squeezing and broadband entanglement \cite{Grimsmo2017,Esposito2021,Fasolo2021}.

Two-mode squeezing in superconducting circuits has been demonstrated in narrow-band Josephson parametric amplifiers based on resonant structures \cite{Eichler2011, Flurin2012,Flurin2015,Menzel2012,Ku2015,Fedorov2016,Fedorov2018b,Westig2017}, semi infinite transmission lines via Dynamical Casimir Effect \cite{Forgues2015,Wilson2011a,Schneider2020} and surface acoustic wave hybrid systems \cite{Andersson2020}.

The use of TWPAs as entanglement and two-mode squeezing sources would have the advantage of a large instantaneous bandwidth and customizable nonlinearities, with potential novel applications for quantum sensing \cite{Degen17,Danilin2021}, quantum enhanced detection \cite{Backes2021}, quantum teleportation with propagating waves \cite{Fedorov2021} and quantum information with continuous variables \cite{Braunstein2005, Weedbrook2012, Hillmann2020}.

For TWPA devices, however, the presence of loss \cite{Houde2019, Zhao2021, Shu2021} and the activation of spurious nonlinear processes, such as harmonics \cite{OBrien2014} and sidebands generation \cite{Peng2021}, have very soon been identified as strong limitations for the observation of squeezing.

In this letter, by using an improved fabrication process \cite{Planat2019b,Planat2020a} and optimizing the device length to mitigate internal loss, we demonstrate generation of vacuum two-mode squeezing in a Josephson TWPA.  

The device is driven with a pump microwave tone at frequency $f_p$ and the generated two-mode quantum state is characterized via repeated measurements of the field quadratures at the signal and idler frequencies $f_s$ and $f_i$, such that $2 f_p = f_s + f_i$ (four wave mixing interaction). 
Given the broadband nature of TWPAs, entangled pairs of signal/idler photons are generated in the entire amplification bandwidth of the device. 
The entanglement between signal and idler modes is quantified by the reconstruction of the two-mode covariance matrix \cite{Olivares2012} and the estimation of the logarithmic negativity \cite{Adesso2005}. 
We obtain non-zero logarithmic negativity and inferred squeezing of the collective quadratures between 1.5 and 2.1 dB below the vacuum limit, for a maximum frequency separation between signal and idler of $400$ MHz, set by the capability of the adopted experimental setup.
In addition, we verify the stability in time of the entanglement generation by studying the statistical distribution of experiments repeated over time scale of hours.
    \begin{figure*}[htb]
        \centering
        \includegraphics[width=0.8\textwidth]{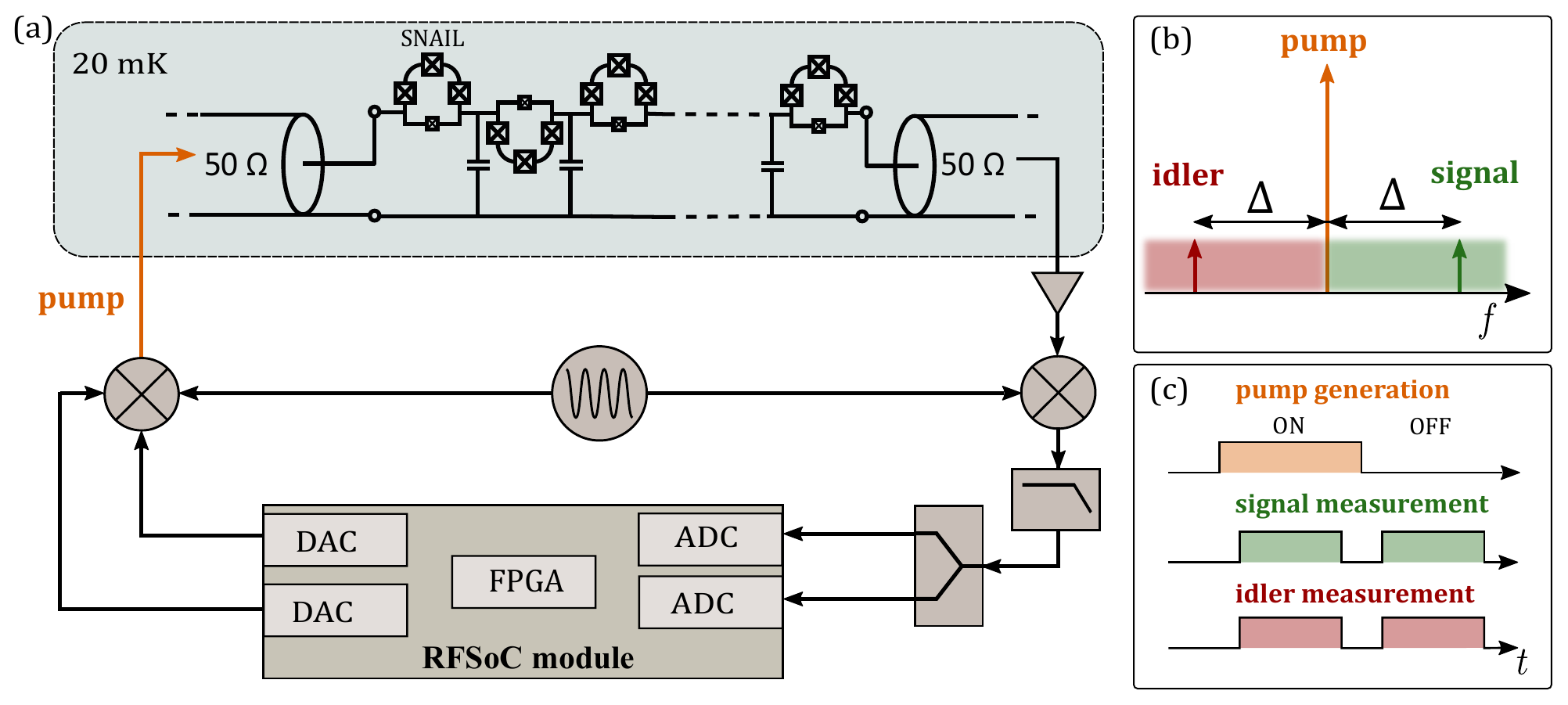}
        \caption{(a) Simplified sketch of the experimental setup for two-mode squeezing generation and detection. The Josephson TWPA device is composed of 250 SNAIL-based unit cells and it is anchored at the 20 mK stage of a dilution refrigerator. The input pump field is obtained via up conversion with an IQ mixer: the I and Q inputs at an intermediate frequency ($f_{\text{pump-IF}} = 290$~MHz) are mixed with a microwave local oscillator (generated with a RF source). The output microwave radiation is amplified, down converted with a mixer, split in two path and finally digitized. The room temperature electronics is composed by an RF System-on-Chip (RFSoC) acquisition board with integrated FPGA (Field Programmable Gate Array), two Digital to Analog Converters (DACs), used to generate the pulsed intermediate frequency inputs for the IQ mixer, and two Analog to Digital Converters (ADCs), used for digitizing the down-converted output radiation. (b) Sketch of the frequency spacing between pump and a pair of signal and idler. (c) Sketch of the control and readout pulse sequence for a single pump on/off generation and measurement; the duration of the readout pulses (acquisition time) is $6 \, \mu\text{s}$.}
        \label{setup_exp}
    \end{figure*}

The adopted device is a Josephson TWPA whose unit cell consists in a superconducting nonlinear asymmetric inductive element (SNAIL) \cite{Frattini2017b} operated in a dilution refrigerator at 20 mK. Such Josephson meta-material has been recently demonstrated and successfully operated as near quantum-limit broadband microwave amplifier \cite{Ranadive2021}. The device used in this experiment is similar to the one presented in \cite{Ranadive2021}, with the only difference of a reduced length of the medium (250 cells) for loss mitigation (see Supplemental Material for details). The device is operated in the 4-wave mixing regime at zero flux. An external magnetic flux threading the SNAIL loops is set by a superconducting coil located alongside the device sample holder to counter stray fields.

When the TWPA is driven with pump frequency $f_{p}$, photon pairs are generated at signal and idler frequencies $f_s$ and $f_i$ such that $f_{s/i} = f_p \pm \Delta$, via a two-mode squeezing interaction \cite{Olivares2012}.
In absence of other frequency conversion or wave-mixing processes and neglecting losses, the input two-mode (signal-idler) vacuum state is transformed at the output of the TWPA into a two-mode squeezed state: $\ket{\text{TMS}}=\hat{S}\ket{0}_{s} \ket{0}_{i}$. The two-mode squeezing operator is:
\begin{equation}
\label{TMS}
     \hat{S} = \text{exp}\left( \xi \hat{a}_{s, \mathrm{in}}^{\dagger}  \hat{a}_{i, \mathrm{in}}^{\dagger} - \xi^{*} \hat{a}_{s, \mathrm{in}}  \hat{a}_{i, \mathrm{in}} \right) \, ,%= \frac{1}{\cosh{r}} \sum_{n} (\tanh{r})^n \ket{n}_s \, \ket{n}_i \,  ,
\end{equation}
where ${\xi = r e^{i \phi}}$ is the squeezing parameter and $\hat{a}_{s, \mathrm{in}}$ ($\hat{a}_{i, \mathrm{in}}$) is the signal (idler) mode operator at the input of the TWPA.%\\

A sketch of the experimental setup and device is shown in Fig \ref{setup_exp} (a). 
The device is driven with a pump tone and the two-mode quantum state generated at the output undergoes an amplification chain, from 20 mK to room temperature, with total system gain $G_{\text{sys}}$. 
At room temperature, a double heterodyne detection scheme allows the detection of signal and idler fields: the output microwave radiation is analogically down-converted to an intermediate frequency, split in two paths and digitized using a multi-channel acquisition board with a sampling clock set to 2 GSamples/s (see Supplemental Material for a detailed description of the setup). Digital down conversion is then performed on board.
% at signal and idler frequencies $f_{s/i} = f_p \pm \Delta$, with detuning $\Delta = 200 $~MHz. 
The resolution bandwidth is defined by the inverse of the acquisition time $\tau = 6 \, \mu\text{s}$.%\\
%%%

The raw measured observables are the signal and idler field quadratures,
    \begin{equation}
        \hat{X}_{s/i} = \frac{1}{2}\left( \hat{A}_{s/i} + \hat{A}^{\dagger}_{s/i} \right) \, , \qquad \hat{P}_{s/i} = \frac{1}{2 i}\left( \hat{A}_{s/i} - \hat{A}^{\dagger}_{s/i} \right),
    \end{equation}
where $\hat{A}^{\dagger}_{s/i}$ and $\hat{A}_{s/i}$ are the creation and annihilation operators for the signal and idler mode such that ${[\hat{A}_j, \hat{A}^{\dagger}_k] = \delta_{j,k}}$.%\\

A single acquisition consists in measuring the raw quadratures with the pump both switched off and switched on. A sketch of the control/readout pulse sequence for a single on/off acquisition is shown in Fig \ref{setup_exp} (c). 

Once the raw quadrature data are measured, they are normalized by the gain of the detection chain at signal and idler frequency respectively: 
$$
x_{s/i} = X_{s/i}/\sqrt{G_{\text{sys}, \, s/i}}  \qquad , \qquad p_{s/i} = P_{s/i}/\sqrt{G_{\text{sys}, \, s/i}}  \, .
$$
%%%%%
% $$
% x_{i} = X_{i}/\sqrt{G_{\text{sys}, i}} \qquad , \qquad p_{i} = P_{i}/\sqrt{G_{\text{sys}, i}} \, .
% $$
The estimation of the system gain, $G_{\text{sys}, \, s/i}$, is obtained by using a shot noise tunnel junction (SNTJ) noise source \cite{Spietz2003,Spietz2006,Chang2016} (see Supplemental Material for details). 

We stress that a careful calibration of $G_{\text{sys}, \, s/i}$ is necessary for a quantitative estimation of the amount of entanglement and squeezing. It should be noticed that here we use an upper bound estimation for $G_{sys}$ (see Supplemental Material). This choice gives a rigorous lower bound estimation for the amount of entanglement and squeezing.
%%%%%  

    \begin{figure*}[htbp]
        \centering
        \includegraphics[width=1.0\textwidth]{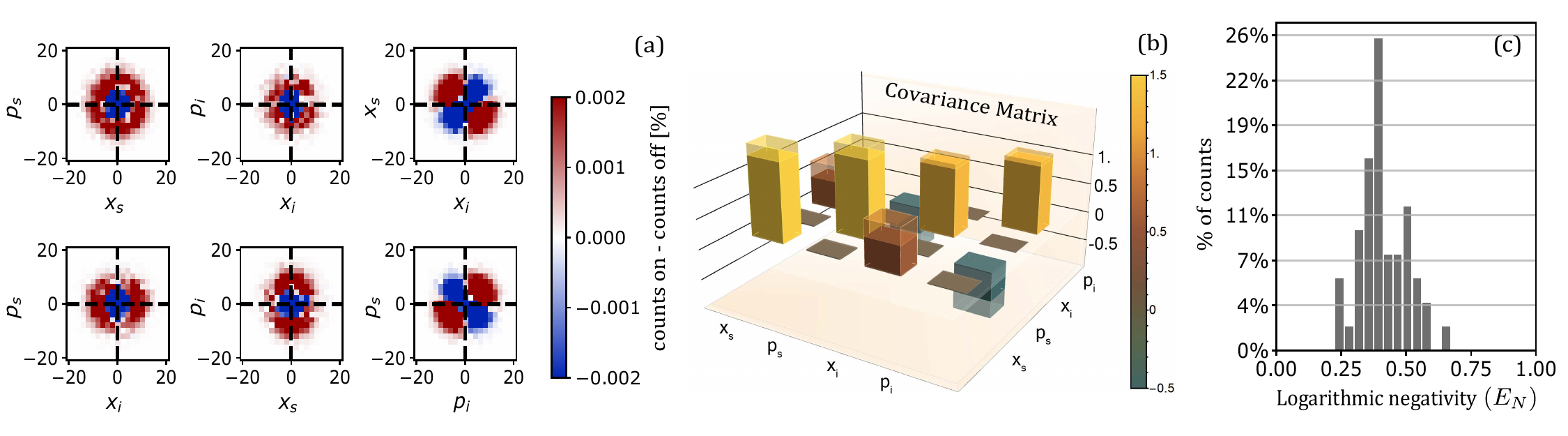}
        \caption{\textbf{} (a) Measured quadrature phase space distribution (2-dimensional differential histogram plots: difference between pump on and pump off histograms) for $10^8$ on/off acquisitions and detuning $\Delta$ = 200~MHz; the shape of the differential quadrature distributions in the ($x_i$, $x_s$) and ($p_i$, $p_s$) phase spaces indicate presence of two-mode correlations.  (b) Reconstructed covariance matrix with uncertainty indicated by shaded regions; (c) Entanglement stability over time: histogram of 50 repeated entanglement measurements, each obtained from a set of $2\times10^6$ repeated on/off quadrature acquisitions.  Pump frequency $f_p = 4.415$ GHz.}
        \label{quadratures}
    \end{figure*}

The same acquisition is repeated for $N_{\text{rep}}$ times, obtaining, for both pump off and on, 6 two-dimensional quadrature histograms (combinations of the four measured quadratures). For $N_{\text{rep}} = 10^8$ the full on/off experiment takes about one hour.\\

Experimentally obtained quadrature histograms for detuning $\Delta$ = 200~MHz and $N_{\text{rep}} = 10^8$ are shown in Fig \ref{quadratures} (a). Each two-dimensional histogram is obtained by subtracting the histogram with pump-off from the histogram with pump-on. 

Starting from the measured quadratures we estimate the covariance matrix \cite{Olivares2012} of the generated bipartite quantum state. This matrix encodes all the quantum properties of the generated state. We compute the covariance matrix from the experimental data for both pump-on and pump-off as:
\begin{equation}
\mathbf{\sigma}_{jk}^{\text{meas}} = 4\left[\frac{1}{2}\braket{R_{j} R_{k} + R_{k} R_{j} } - \braket{R_{j}} \braket{ R_{k}}\right], 
\end{equation}
%%%
where $\mathbf{R} = (x_s, p_s, x_i, p_i)$. 
%With this convention, the covariance matrix for the vacuum state reduces to the identity matrix. \\
The covariance matrix of the two-mode quantum state generated at the output of the TWPA can be inferred by subtracting the pump-off noise background as follows:  \cite{Flurin2015}:
    \begin{equation}
        \mathbf{\sigma} =  \mathbf{\sigma}^{\mathrm{meas, on}} -\mathbf{\sigma}^{\mathrm{meas, off}} +  \mathbb{1}_{4} \, ,\\
    \end{equation}
where, with the adopted convention, the two-mode vacuum state covariance matrix corresponds to
the unit matrix.

The inferred covariance matrix is depicted in Fig \ref{quadratures} (b). The presence of not vanishing off-diagonal elements indicates two-mode squeezing correlations between signal and idler, verifying a necessary condition for the demonstration of entanglement generation. %\\
% The covariance matrix encodes the maximum information about a bipartite Gaussian state \cite{Weedbrook2012}. 

A quantitative entanglement estimation is provided by the logarithmic negativity $E_{\mathcal{N}}$ \cite{Adesso2005} defined as $E_{\mathcal{N}} = \mathrm{Max}\left[ -\ln{ ( \nu_{-})}, 0\right]$, where $\nu_{-}$ is the minimum symplectic eigenvalue of the partially trasposed covariance matrix (see Supplemental Material). Following the partial positive transpose (PPT) criterion \cite{Horodecki1997}, the logarithmic negativity being positive ($E_{\mathcal{N}} > 0$) is a sufficient condition to demonstrate that the state is entangled. %\\ 

For the data shown in Fig \ref{quadratures} we measure a logarithmic negativity $E_{\mathcal{N}} = 0.4\pm 0.1$, which proves entanglement generation, i.e. quantum nature of observed correlations. 

The pump power is optimized to get the maximum entanglement (maximum logarithmic negativity). This is reached for a linear gain $G~=~1.30 \pm 0.05$, corresponding to a squeezing parameter $r = \text{arcosh}(\sqrt{G}) = 0.52 \pm 0.04$. For higher pump powers, that is higher TWPA gain G, %the activation of spurious wave mixing processes limits the exploration of squeezing. A detailed study of the high pump power regime is beyond the scope of this work.\\
we observe a decrease of the logarithmic negativity (see Supplemental Material). Similar behavior has been observed in resonant Josephson parametric amplifiers \cite{Boutin17} and semi-infinite transmission lines \cite{Schneider2020}, due to the activation of spurious higher order and wave-mixing processes. In TWPAs, the role of such spurious processes on limiting the amount of generated squeezing is currently an open question. Further explorations of the high pump power regime is beyond the scope of this letter.\\

%%%%

The stability of the generated entanglement over time is verified performing 50 repetitions of the same experiment (each consisting of a set of $2\times 10^6$ quadrature acquisitions) over a total time of roughly one hour. In Fig \ref{quadratures} (c) we show the obtained histogram distribution of the logarithmic negativity. 
This result demonstrates that the generated entanglement is stable for repeated measurements over time scale of hours.\\

A largely adopted second criterion to verify the non-classicality of the generated bipartite state, consists in estimating the variances of the collective quadratures, 
    \begin{equation}
        \hat{x}_{+} = \left( \hat{x}_{s} + \hat{x}_{ i} \right)\, , \qquad \hat{p}_{+} = \left( \hat{p}_{s} + \hat{p}_{ i} \right)\, , 
    \end{equation}
 and compare them with the vacuum noise \cite{Eichler2011,Flurin2012,Houde2019}.%\\

Considering a squeezing parameter amplitude $r$ and accounting for average TWPA loss, $ 0 \leq \bar{\epsilon} \,  \leq 1$, with a simple lumped-element beam splitter model (see Supplemental Material), the variance of the collective quadrature can be expressed as \cite{Houde2019}: 
\begin{equation}
    \label{var_loss}
    \braket{x_{+}^2} = \frac{1}{2} \, \left[ \bar{\epsilon} + (1 - \bar{\epsilon} ) e^{-2r}   \right]  \, .
\end{equation}
For a non-classical state, the latter is expected to be lower than the vacuum quantum noise giving non-zero squeezing, defined as $\text{Sq}_{+} = 10\log\left(\frac{\braket{x_{+}^2} }{0.5}\right)$.%\\

For the data shown in Fig \ref{quadratures}, we obtain a lower bound on the squeezing value: $\text{Sq}_{+}=-1.6 \pm 0.5$~dB below the vacuum limit. 

In the adopted pump power regime, the obtained squeezing value is limited by the effect of internal loss present in the device: for fixed TWPA gain, a device with higher loss (composed of a higher number of unit cells) will generate a reduced amount of squeezing. We designed the length of the TWPA device, 250 unit cells, to mitigate internal loss and foster the squeezing generation for a fixed gain value (see Supplemental Material). \\%The activation of higher order non-linear processes limits the exploration of squeezing at larger pump powers. A detailed study of the high pump power regime is beyond the scope of this work. \\

\begin{figure}[htbp]
    \centering
    \includegraphics[width=0.5\textwidth]{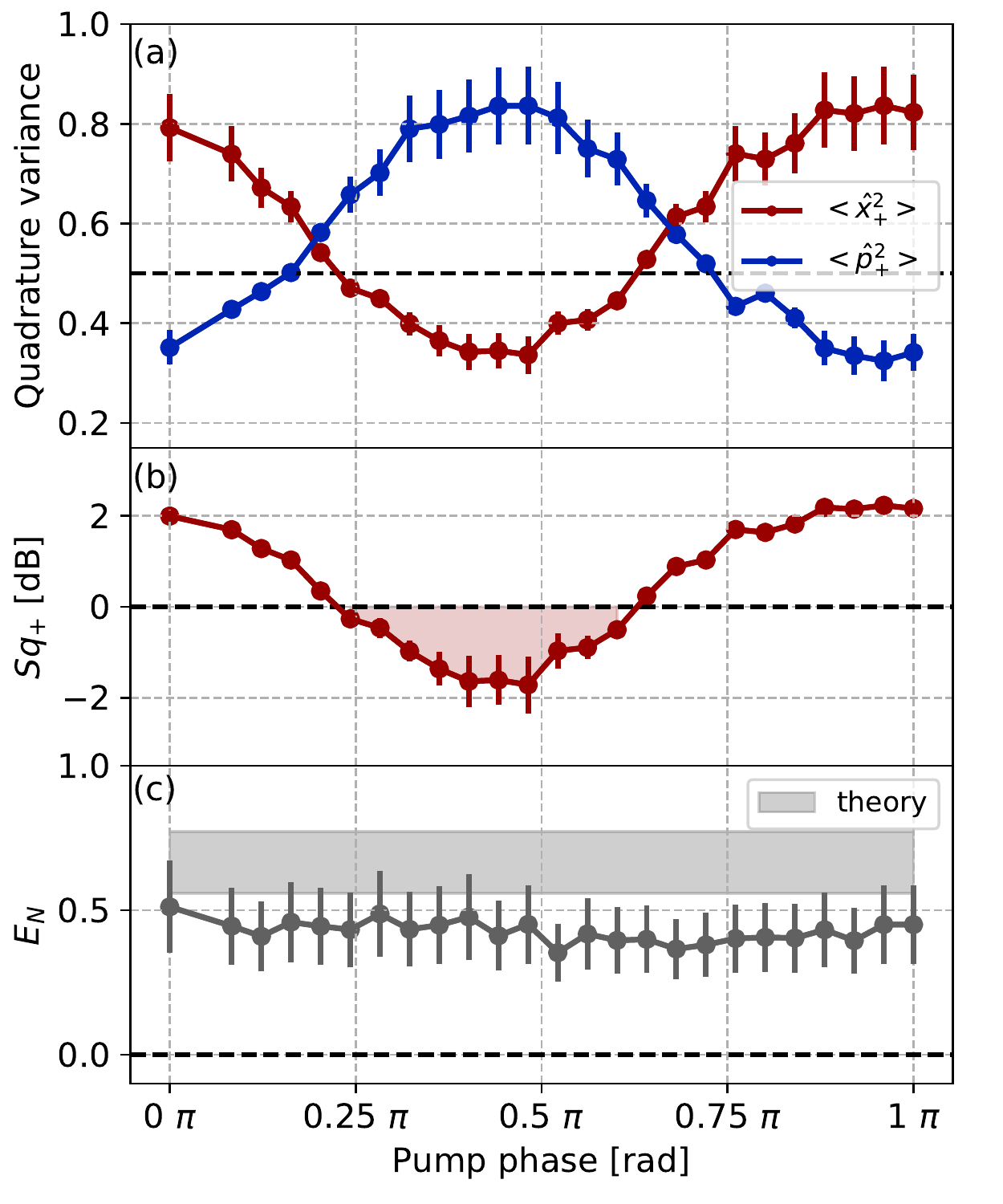}
    \caption{\textbf{}Collective quadrature variance (a), squeezing (b), and logarithmic negativity (c) as a function of the input pump phase. Pump frequency $f_p = 4.415$ GHz, detuning $\Delta$ = 200~MHz, number of repeated measurements for each phase $N_{\text{rep}} = 10^7$. Dashed lines indicate the vacuum state reference. Squeezing below the vacuum limit is highlighted with a shadow area in panel (b). The gray shadow area in panel (c) indicates the logarithmic negativity predicted by a simple lumped-element model which takes into account the TWPA gain and the estimated losses (see Supplemental Material).}
    \label{squeezing_phase}
\end{figure}

In Fig \ref{squeezing_phase} we show the pump phase-dependence of the generated two-mode squeezing.
By varying the pump input phase, we observe the expected periodicity of the collective quadrature variances, panel (a), and consequently of the squeezing $\text{Sq}_{+}$, panel (b). In panel (c), the experimentally obtained logarithmic negativity $E_{\mathcal{N}}$ is reported. We stress that this is a conservative lower bound estimation (an upper bound estimation for $G_{sys}$ is considered, giving a lower bound estimation of the amount of entanglement). The gray shadow area indicates the logarithmic negativiy predicted by a simple two-mode squeezing model that considers the measured TWPA gain and estimated device losses (see Supplemental Material for details).\\

A key advantage in generating two-mode squeezing with TWPAs is their broadband nature: when the pump drives the device, all the pairs of signal and idler modes within the TWPA bandwidth (several GHz \cite{Esposito2021}) get entangled. 
To quantify the broadband nature of the generated entangled states, we estimate the rate of entangled bits generation by calculating the entropy formation, $E_{F} = c_{+} \log_2{c_{+}} - c_{-} \log_2{c_{-}}$, with $c_{\pm} = (\delta^{-1/2} \pm \delta^{1/2})^2/4$ and $\delta = 2^{E_{\mathcal{N}}}$ \cite{Giedke2003,Flurin2012}, and multiplying it by the squeezing bandwidth $2\Delta$.  
Despite the fact that bandwidth of our device is in the GHz range, we could only check entanglement between signal and idler separated by a maximum of $2 \Delta = 400$~MHz, because of limitations of the adopted acquisition system. We use such bandwidth to obtain a lower bound of $53 \pm 20$ Mebit/s (mega entangled bits per second) on the rate of entanglement generation, which is already comparable with previous state of the art (Supplemental Material of \cite{Schneider2020}).\\ 

\begin{figure}[htbp]
    \centering
    \includegraphics[width=0.5\textwidth]{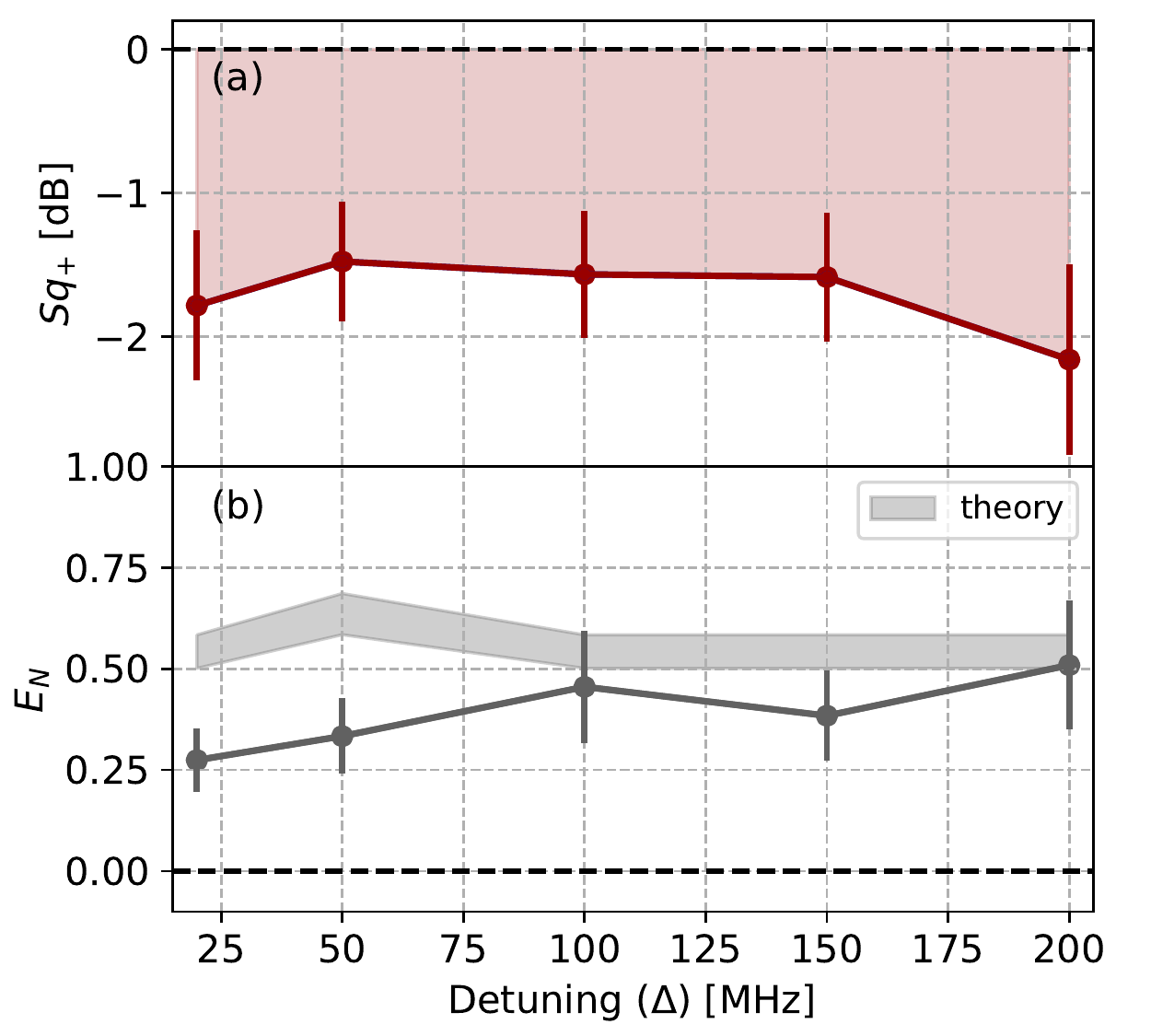}
    \caption{\textbf{} Two-mode squeezing (a) and logarithmic negativity (b) as a function of detuning $\Delta$.  Pump frequency: $f_p = 4.415$ GHz, number of repeated measurements for each $\Delta$: $N_{\text{rep}} = 10^7$. Dashed lines indicate the vacuum state reference. The gray shadow area in panel (b) indicates the logarithmic negativity predicted by a simple lumped-element model which takes into account the TWPA gain and the estimated loss (see Supplemental Material).}
    \label{sweep_delta}
\end{figure}

Finally, to further investigate the entanglement generation within the available frequency bandwidth, we vary the detuning $\Delta$ between the pump frequency and the signal/idler frequencies. For each value of the detuning we measure the logarithmic negativity and the squeezing. The results are shown in Fig \ref{sweep_delta}. 
We observe entanglement and two-mode squeezing in the entire investigated frequency region. This result further demonstrates the truly broadband nature of squeezing generation in the device under study. We again report the comparison between our lower bound estimation of the logarithmic negativity and the prediction of a simple lumped-element model which takes into account measured TWPA gain and internal loss. \\

In conclusion, we demonstrated two-mode squeezing generation in a traveling wave parametric amplifier. This result proves that TWPAs can be effectively used as sources of microwave radiation. We obtained broadband squeezing of between 1.5 and 2.1 dB for signal and idler modes separated by up to 400 MHz. 
This first result can be further improved via a deeper understanding of loss mechanisms and higher order non-linear effects in TWPAs. Another interesting direction will be to take advantage of flexibility offered by TWPAs to explore different wave-mixing processes and tailored nonlinearities. 

On a broad prospective, our findings experimentally demonstrate the potential of TWPA devices beyond amplification, paving the way to their application in the context of microwave photonics, quantum sensing and quantum information with continuous variables. \\

The data supporting the findings presented in this work are openly available at https://doi.org/10.5281/zenodo.5217996.\\

%%%% Supplemental Material Refrences
See Supplemental Material for details on the experimental setup, models and calibration procedures, which includes Refs. \cite{Lang2014, Malnou2021, dir_coupler, cryo-switch}.

%% Commercial Product Disclaimer
Certain commercial equipment, instruments, or materials are identified in this paper in order to specify the experimental procedure adequately.  Such identification is not intended to imply recommendation or endorsement, nor is it intended to imply that the materials or equipment identified are necessarily the best available for the purpose.\\

% Acknowledgements
This work is supported by the European Union’s Horizon 2020 research and innovation program under grant agreement no. 899561.  M.E. acknowledges the European Union’s Horizon 2020 research and innovation program under the Marie Sklodowska Curie (grant agreement no. MSCA-IF-835791). A.R. acknowledges the European Union’s Horizon 2020 research and innovation program under the Marie Sklodowska Curie grant agreement No 754303 and the ’Investissements d’avenir’ (ANR-15-IDEX-02) programs of the French National Research Agency. The sample was fabricated in the clean room facility
of Institute Neel, Grenoble, we sincerely thank all the clean room staff for help with fabrication of the devices. We would like to acknowledge E. Eyraud for his extensive help in the installation and maintenance of the cryogenic setup and J. Minet for significant help in programming the high speed pulse generation and data acquisition setup. We also thank J. Jarreau, L. Del Rey for their support with the experimental equipment.
We are grateful to B. Boulanger, A. Metelmann and S. Böhling for insightful discussions regarding this project. We thank the members of the superconducting circuits group at Neel Institute for helpful discussions. We sincerely thank M. Malnou and J. D. Teufel for their careful reading of the manuscript.\\

% \bibliographystyle{ieee}
%\bibliography{zPaper_two_mode_squeezing.bib}% Produces the 
% \bibliography{C:/Users/martina.esposito/Dropbox/work/literature/bib_files/library.bib}% Produces the bibliography via BibTeX.

%

\end{document}

\preprint{APS/123-QED}

\title{Supplemental Material: Observation of two-mode squeezing in a traveling wave parametric amplifier}
    \author{Martina Esposito}
    \thanks{These authors contributed equally to the work.}
    \affiliation{Univ. Grenoble Alpes, CNRS, Grenoble INP, Institut N\'eel, 38000 Grenoble, France}
    \affiliation{CNR-SPIN Complesso di Monte S. Angelo, via Cintia, Napoli, 80126, Italy}
    \author{Arpit Ranadive} 
    \thanks{These authors contributed equally to the work.}
    \affiliation{Univ. Grenoble Alpes, CNRS, Grenoble INP, Institut N\'eel, 38000 Grenoble, France}
    \author{Luca Planat}
    \affiliation{Univ. Grenoble Alpes, CNRS, Grenoble INP, Institut N\'eel, 38000 Grenoble, France}
    \author{Sébastien Léger}
    \affiliation{Univ. Grenoble Alpes, CNRS, Grenoble INP, Institut N\'eel, 38000 Grenoble, France}
    \author{Dorian Fraudet}
    \affiliation{Univ. Grenoble Alpes, CNRS, Grenoble INP, Institut N\'eel, 38000 Grenoble, France}
    \author{Vincent Jouanny}
    \affiliation{Univ. Grenoble Alpes, CNRS, Grenoble INP, Institut N\'eel, 38000 Grenoble, France}
    \author{Olivier Buisson}
    \affiliation{Univ. Grenoble Alpes, CNRS, Grenoble INP, Institut N\'eel, 38000 Grenoble, France}
    \author{Wiebke Guichard}
    \affiliation{Univ. Grenoble Alpes, CNRS, Grenoble INP, Institut N\'eel, 38000 Grenoble, France}
    \author{C\'ecile Naud}
    \affiliation{Univ. Grenoble Alpes, CNRS, Grenoble INP, Institut N\'eel, 38000 Grenoble, France}
    \author{José Aumentado}
    \affiliation{National Institute of Standards and Technology, Boulder, Colorado, 80305, USA}
    \author{Florent Lecocq}
    \affiliation{National Institute of Standards and Technology, Boulder, Colorado, 80305, USA}
    \author{Nicolas Roch}
    \affiliation{Univ. Grenoble Alpes, CNRS, Grenoble INP, Institut N\'eel, 38000 Grenoble, France}

\date{\today}% It is always \today, today,

\maketitle

%\tableofcontents
%%%%%%%%%%%%%%%%%%%%%%%%%%%%%%%%%%%%%%%%%%%%%%%%%%%%%%%%%%%%%%%%%%%%%
\section{\label{intro}Experimental setup}
%%%%%%%%%%%%%%%%%%%%%%%%%%%%%%%%%%%%%%%%%%%%%%%%%%%%%%%%%%%%%%%%%%%%%
\begin{figure}[hbtp]
    \centering
    \includegraphics[width=0.9\columnwidth]{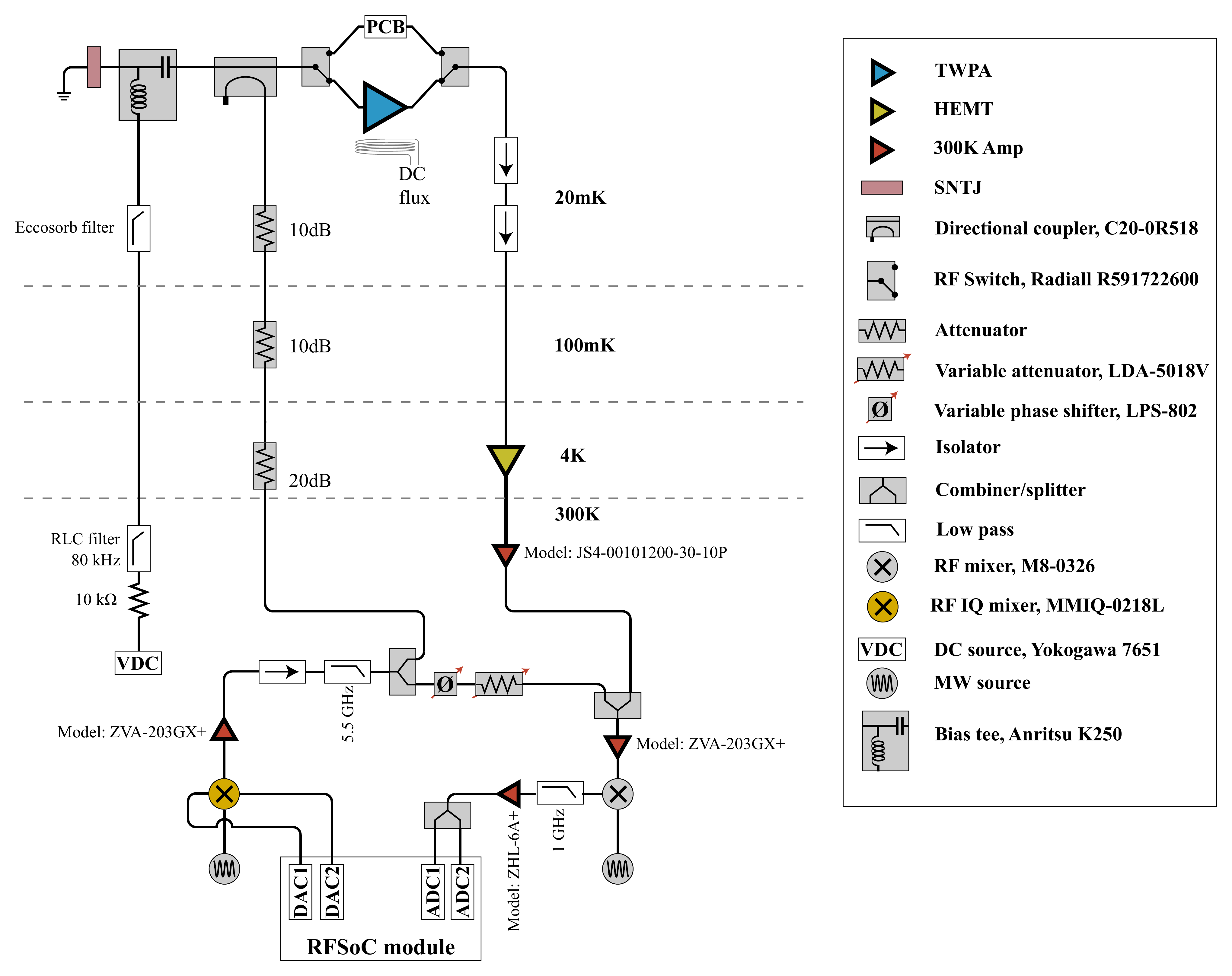}
    \caption{\textbf{Sketch of the experimental setup for the measurement of two-mode squeezing.}}
    \label{setup}
\end{figure}
The scheme of the experimental setup is shown in Fig \ref{setup}. 
The TWPA device is anchored to the coldest plate of a dilution refrigerator, at 20 mK. The device consists of a chain of 250 superconducting nonlinear asymmetric elements (SNAILs) \cite{Frattini2017b}. Details on device fabrication and design can be found in \cite{Ranadive2021}. 
We use two cryogenic microwave switches at 20 mK to switch between the TWPA device and a reference device which consists in a PCB CPW transmission line. We use the PCB (reference device) to perform the calibration of the system gain (total gain of the detection chain) via a calibrated noise source, shot noise tunnel junction (SNTJ), as explained in section \ref{SNTJ}. 
A superconducting coil is placed alongside the device holder allowing to control the magnetic flux threading the SNAILs.

At room temperature, the pump tone is generated via up conversion by using an IQ mixer: the I and Q pulsed inputs at an intermediate frequency ($f_{\text{pump-IF}} = 290$ MHz) are generated by two digital to analog converters (DACs), while the local oscillator input is generated with a RF source. 
For the readout, a two step demodulation of the microwave output radiation is performed \cite{Lang2014}: the output is down-converted with a mixer to an intermediate frequency band (with $f_{\text{pump-IF}}$ being the central frequency), split in two paths, digitized with two analog to digital converters (ADCs), and finally a digital down conversion is performed from the intermediate frequency to DC. We need to split the output in two paths because of a limitation of our present FPGA firmware (it does not allow us to use the data coming from a single ADC to digitally down convert two different frequencies to DC). With a future upgrade of the FPGA firmware, using just a single ADC, one will be able to acquire quadratures at different frequencies within the intermediate frequency band.

The room temperature electronics is composed by a multi-channel Xilinx RF System-on-Chip (RFSoC) acquisition board which includes a Field Programmable Gate Array (FPGA), 8 Digital to Analog Converters (DACs) and 8 Analog to Digital Converters (ADCs) with 2 GS/s sampling rate. 
The hardware and software architectures of the acquisition board have been developed by the electronic engineering group at the Neel institute in Grenoble. 

The sensitivity of the quadrature measurement is dictated by the sensitivity of our heterodyne detection, which is currently limited by the noise performance of the first amplifier (HEMT) in the measurement setup. This sensitivity is 2.3 in units of square root of number of photons.

An interferometric cancellation of the output pump field is performed at room temperature using a variable attenuator and a variable phase-shifter. Pump field cancellation at room temperature is used to avoid saturation of the acquisition board due to the intense transmitted pump field.

The same setup is used also for the measurement of the TWPA gain $G$. In that case, a weak tone at the signal frequency ($f_s = f_p + \Delta$) is injected into the device together with the strong tone at the pump frequency $f_p$. To do so, the I and Q pulsed inputs (digitally generated with the DACs) contains both a component at the pump intermediate frequency ($f_{\text{pump-IF}} = 290$ MHz) and a weaker component at the signal intermediate frequency ($f_{\text{signal-IF}} = f_{\text{pump-IF}}~-~\Delta$). 
The output power at the signal frequency is measured (by using the two step demodulation explained earlier) for pump tone off and on, allowing to extract the TWPA gain. %The error on the estimated TWPA gain is the standard deviation over $10^5$ repeated measurements.   

%%%%%%%%%%%%%%%%%%%%%%%%%%%%%%%%%%%%%%%%%%%%%%%%%%%%%%%%%%%%%%%%%%%%%
\section{\label{device}Device}
%%%%%%%%%%%%%%%%%%%%%%%%%%%%%%%%%%%%%%%%%%%%%%%%%%%%%%%%%%%%%%%%%%%%%
The device is a Josephson-junction meta-material fabricated with standard double-angle Aluminum evaporation techniques, followed by dielectric and top-ground deposition \cite{Planat2019b}. Design and fabrication of the device used in this experiment are identical to the one presented in \cite{Ranadive2021}. The unit cell consists in a superconducting loop with 3 big Josephson junctions (critical current $I_0$) and one small Josephson junction (critical current $r I_0$). The adopted device is composed by 250 unit cells. 
The values of $r = 0.05$ and $I_0 = 1.47 \mu$A are obtained from the device characterization in linear regime (with no pump applied).

The Josephson capacitance per unit cell is $C_J = 31$ fF, while the ground capacitance per unit cell is $C_g = 550$ fF; they have been also estimated from the linear characterization of the device \cite{Ranadive2021}.\\

TWPA losses are estimated by measuring the transmission ($S_{21}$ scattering parameter) of the TWPA in the linear regime (no pump tone applied) with a Vector Network Analyzer (VNA). The $S_{21}$ transmission measurement is performed both through the TWPA device and through the reference device (dubbed 'PCB' in Fig. \ref{setup}), consisting in a copper PCB CPW transmission line of identical size of the TWPA chip wire-bonded in a sample box identical to the one of the TWPA chip. We can switch from the TWPA device to the PCB within the same cool-down by means of two cryogenic microwave switches (see Fig. \ref{setup}). The cables used to connect the TWPA and the PCB to the switches are nominally identical. 
% \red{In Fig \ref{sample_box} we show a photo of the PCB and TWPA packaging.}
% \begin{figure}[hbtp]
%     \centering
%     \includegraphics[width=0.3\textwidth]{sample_box_V1}
%     \caption{Picture of PCB (left) and TWPA (right) identical sample boxes.}
%     \label{sample_box}
% \end{figure}
By subtracting the $S_{21}$ PCB transmission from the $S_{21}$ TWPA transmission we can estimate the TWPA losses. Measured TWPA losses are found to be between $0.5$ and $1$ dB in the 4-6 GHz frequency region, primary arising from dielectric losses \cite{Ranadive2021}. We estimate loss tangent (tan $\delta_0$) to be approximately $1.6 \times 10^{-3}$ .

%%%%%%%%%%%%%%%%%%%%%%%%%%%%%%%%%%%%%%%%%%%%%%%%%%%%%%%%%%%%%%%%%%%
\section{Two mode squeezing generation}
%%%%%%%%%%%%%%%%%%%%%%%%%%%%%%%%%%%%%%%%%%%%%%%%%%%%%%%%%%%%%%%%%%%%
Signal and idler mode operators under the action of a two-mode squeezing interaction evolve as follows:
\begin{equation}
    \hat{a}_{s, \mathrm{out}} = \hat{S}^{\dagger} \, \hat{a}_{s, \mathrm{in}} \, \hat{S} = \cosh{r} \, \hat{a}_{s, \mathrm{in}} + e^{i \phi} \sinh{r} \, \hat{a}_{i, \mathrm{in}}^{\dagger} \, ,
\end{equation}
\begin{equation}
    \hat{a}_{i, \mathrm{out}} = \hat{S}^{\dagger} \, \hat{a}_{i, \mathrm{in}} \, \hat{S} = \cosh{r} \, \hat{a}_{i, \mathrm{in}} + e^{i \phi} \sinh{r} \, \hat{a}_{s, \mathrm{in}}^{\dagger} \, ,
\end{equation}
where $\hat{S} = \mathrm{exp}\left( \xi \hat{a}_{s, \mathrm{in}}^{\dagger}  \hat{a}_{i, \mathrm{in}}^{\dagger} - \xi^{*} \hat{a}_{s, \mathrm{in}}  \hat{a}_{i, \mathrm{in}} \right)$ is the two-mode squeezing operator and ${\xi = r e^{i \phi}}$ is the complex  squeezing parameter.

The squeezing amplitude $r$ is related to the TWPA gain $G$, assumed to be identical for signal and idler, $G = \cosh^2{r}$, while the phase $\phi$ depends on the pump field input phase.
%%%%

In absence of other frequency conversion or wave-mixing processes and neglecting losses, the TWPA scattering matrix is the following:
\begin{equation}
\begin{pmatrix}
\cosh{r}  & 0 & \sinh{r} \cos{\phi} & \sinh{r} \sin{\phi}\\
0 & \cosh{r}  & -\sinh{r} \sin{\phi} & \sinh{r}\cos{\phi} \\
\sinh{r} \cos{\phi} & \sinh{r} \sin{\phi} & \cosh{r} & 0\\
-\sinh{r} \sin{\phi} & \sinh{r} \cos{\phi}  & 0 & \cosh{r} 
\end{pmatrix},
\end{equation}
which maps the input vector $(\hat{a}_{s, \mathrm{in}}, \hat{a}_{s, \mathrm{in}}^{\dagger}, \hat{a}_{i, \mathrm{in}}, \hat{a}_{i, \mathrm{in}}^{\dagger})^{T}$ to the corresponding output one.%\\

When the TWPA is pumped (pump field is on), the input two-mode (signal-idler) vacuum state is converted into a two-mode vacuum squeezed state: 
\begin{equation}
    \ket{\text{TMS}}=\hat{S}\ket{0}_{s} \ket{0}_{i} = \frac{1}{\cosh{r}} \sum_{n} (\tanh{r})^n \ket{n}_s \, \ket{n}_i \,  .
\end{equation}
This is a highly entangled state given by the superposition of twin Fock states at signal and idler frequency. \\

% %%%%%%%%%%%%%%%%%%%%%%%%%%%%%%%%%%%%%%%%%%%%%%%%%%%%%%%%%%%%%%%%%%%
% \subsection{\label{losses}Treatment of TWPA losses with beam splitter model}
% %%%%%%%%%%%%%%%%%%%%%%%%%%%%%%%%%%%%%%%%%%%%%%%%%%%%%%%%%%%%%%%%%%%
Internal losses in the TWPA device, can degrade the two-mode squeezed state. A simple model for losses in TWPAs consists in considering the action of two independent beam splitter operators, with transmission coefficients $\eta_{s/i}$, acting after the squeezing operator on the signal and idler fields respectively \cite{Houde2019}. 
In absence of losses, the beam splitter transmits the entire field ($\eta = 1$), while in presence of losses, the beam splitter transmits with efficiency $\eta<1$ \cite{Flurin2012, Houde2019}. %\\

The mode operators after the beam splitter evolution are the following:
    %%%%
    \begin{equation}
        \hat{a}_{s} =  \sqrt{\eta_s} \, \hat{a}_{s, \mathrm{out}} + \sqrt{1 - \eta_s} \, \hat{a}_{s, \mathrm{th}} \, ,
    \end{equation}
    \begin{equation}
        \hat{a}_{i} = \sqrt{\eta_i} \, \hat{a}_{i, \mathrm{out}} + \sqrt{1 - \eta_i} \, \hat{a}_{i, \mathrm{th}} \, ,
    \end{equation}
    %%%%%
where $\eta_{s/i}$ are the transmission rate of the signal and idler through the beam splitter, and $\hat{a}_{s, \mathrm{th}} $ and $\hat{a}_{i, \mathrm{th}}$ are the ladder operator of the thermal states (noise modes) at the second input of each beam splitter. Since losses can be different for signal and idler \cite{Houde2019}, in general $\eta_s \neq \eta_i$.%\\

\begin{figure*}[htb]
    \centering
    \includegraphics[width=0.8\textwidth]{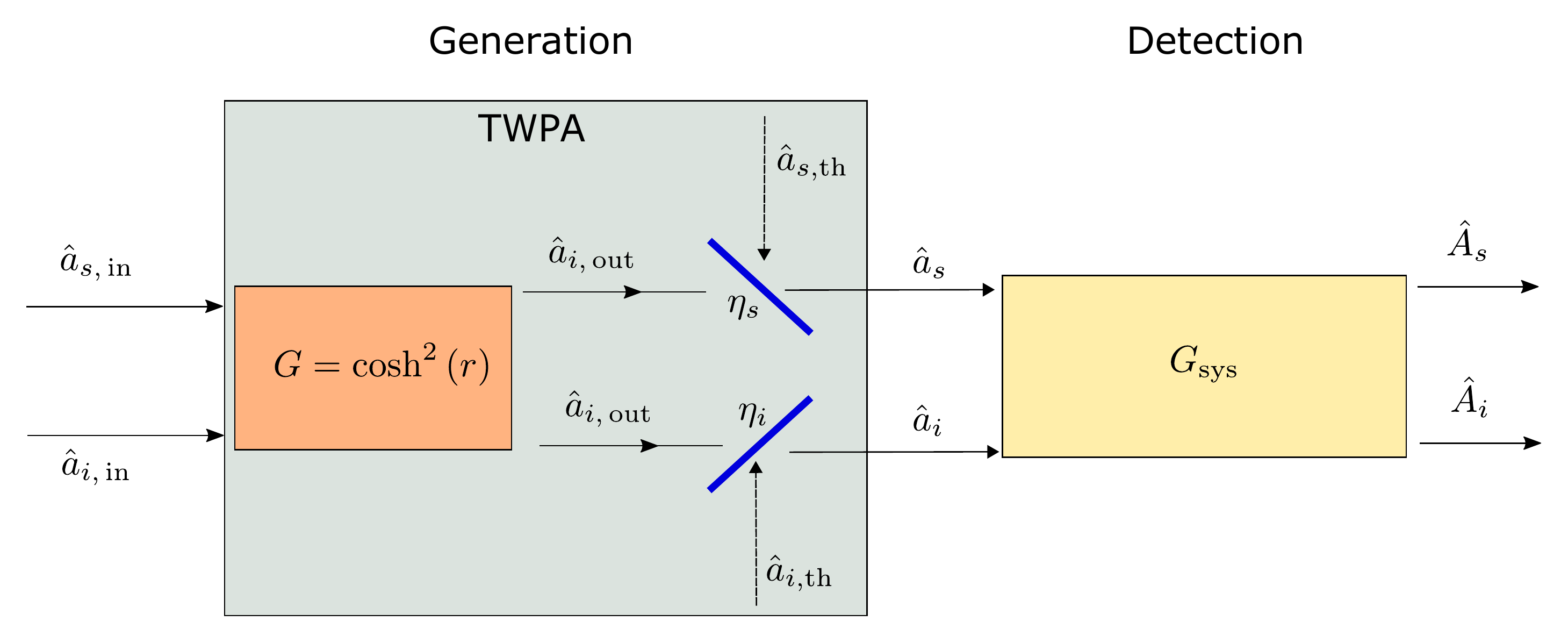}
    \caption{Sketch of two-mode squeezing generation and detection.}
    \label{setup_theory}
\end{figure*}

A sketch of such simple lumped-element model \cite{Houde2019} for generation of two-mode squeezing in a TWPA is shown in Fig \ref{setup_theory}. 
    %%%%%%%%%
   \\
%%%%%%%%%%
The quadrature operators,  $\hat{x}_s$ and $\hat{p}_s$ for the signal field, and $\hat{x}_i$ and $\hat{p}_i$ for the idler field, are defined as follows \cite{Flurin2015, Eichler2011}:
%%%%NOTE: we use the same definition as in \cite{Flurin2015, Eichler2011}
    \begin{equation}
        \hat{x}_{s, i} = \frac{1}{2}\left( \hat{a}_{s, i} + \hat{a}^{\dagger}_{s, i} \right) \, , \qquad \hat{p}_{s, i} = \frac{1}{2 i}\left( \hat{a}_{s, i} - \hat{a}^{\dagger}_{s, i} \right).
    \end{equation}
%%%%%%%%%%%%%%%
The generated two-mode state at the output of the TWPA is a Gaussian state and as such it is completely characterized by its covariance matrix $\mathbf{\sigma}$ \cite{Olivares2012}:
    \begin{equation}
    \mathbf{\sigma}_{jk} = 4\left[\frac{1}{2}\braket{R_{j} R_{k} + R_{k} R_{j} } - \braket{R_{j}} \braket{ R_{k}}\right], 
    \end{equation}
where $\mathbf{R} = (\hat{x}_s, \hat{p}_s, \hat{x}_i, \hat{p}_i)$.
%%%%
The expression for the covariance matrix for a two-mode squeezed state with squeezing parameter $\xi = r e^{i \phi}$ and considering losses with the beam splitter model \cite{Flurin2012, Houde2019} is:
    \begin{widetext}
    \begin{equation}
    \mathbf{\sigma}_{\text{TMS}} = 
    \begin{pmatrix}
     \eta_s \cosh{(2 r)} + (1 - \eta_s) & 0 & \sqrt{\eta_s \, \eta_i}\sinh{(2 r)} \cos{\phi}& \sqrt{\eta_s \, \eta_i}\sinh{(2 r)} \sin{\phi}\\
    0 & \eta_s \cosh{(2 r)} + (1 - \eta_s)  & \sqrt{\eta_s \, \eta_i}\sinh{(2 r)} \sin{\phi} & -\sqrt{\eta_s \, \eta_i} \sinh{(2 r)}\cos{\phi} \\
    \sqrt{\eta_s \, \eta_i} \sinh{(2 r)} \cos{\phi} & \sqrt{\eta_s \, \eta_i}\sinh{(2 r)} \sin{\phi} & \eta_i  \cosh{(2 r)} + (1 - \eta_i) & 0\\
    \sqrt{\eta_s \, \eta_i}\sinh{(2 r)} \sin{\phi} & - \sqrt{\eta_s \, \eta_i} \sinh{(2 r)} \cos{\phi}  & 0 & \eta_i  \cosh{(2 r)} + (1 - \eta_i)
    \end{pmatrix}.
    \label{CM2_complete}
    \end{equation}
    \end{widetext}
%%%%%%%%%%%%%%%%%%%%%%%%%%%%%%%%%%%%%%%%
One can notice that with the convention adopted here, the two-mode vacuum state covariance matrix corresponds to the unit matrix $\mathbb{1}_{4}$. The covariance matrix in \eqref{CM2_complete} is used to estimate the expected logarithmic negativity in the main text (shadow area in Fig 3 and Fig 4), using the measured device gain and estimated losses. In Fig 4, small variations of the model at different detuning values are due to TWPA gain wiggles in the explored frequency band.
%%%%%%%%%%%%%%%%%%%%%%%%%%%%

%%%%%%%%%%%%%%%%%%%%%%%%%%%%%%%%%%%%%%%%%%%%%%%%%%%%%%%%%%%%%%%%%%%
\subsection{Entanglement quantification}
%%%%%%%%%%%%%%%%%%%%%%%%%%%%%%%%%%%%%%%%%%%%%%%%%%%%%%%%%%%%%%%%%%%
It is useful to decompose the two-mode covariance matrix in four 2 X 2 block matrices. The diagonal block matrices represent respectively the signal and idler single-mode covariance matrix, while the off diagonal blocks describe the correlations between the two-modes:
\begin{equation}
     \mathbf{\sigma} =\begin{pmatrix}
A & C \\
C^{T} & B  
\end{pmatrix} .
\label{CM1}
\end{equation}
%%
%%%
We can define the symplectic eigenvalue $\nu_{-}$ as follows \cite{Adesso2005}:%, Asjad2013}:
\begin{equation}
 \nu_{-} = \sqrt{\frac{\Delta{ \mathbf{\sigma} } - \sqrt{(\Delta \mathbf{\sigma} )^2 - 4 \, \mathrm{det}  \mathbf{\sigma} }}{2}}
\end{equation}
where
\begin{equation}
\Delta \mathbf{\sigma} = \mathrm{det} A  + \mathrm{det} B - 2\, \mathrm{det} C \, .
\end{equation}
%The Heisemberg incertanty relation in terms of the symplectic eigenvalue reads $\nu_{-} \ge 1$ \cite{Adesso2005}. 
This quantity encodes the entanglement characterization of the two-mode Gaussian state. In particular, using the PPT criterion \cite{Horodecki1997}, a necessary and sufficient condition for entanglement is 
$$
\nu_{-} < 1 \, .
$$
A typical quantitative measure of entanglement for a bipartite state is the logarithmic negativity $E_{\mathcal{N}}$ \cite{Adesso2005} defined as follows:
\begin{equation}
    E_{\mathcal{N}} = \mathrm{Max}\left[ -\ln{ ( \nu_{-})}, 0\right] \, .
\end{equation}
The two-modes are entangled when $E_{\mathcal{N}} > 0$.
For a two-mode squeezed state with squeezing parameter amplitude $r$ and accounting for loss with the beam splitter model, the logarithmic negativity is expected to be \cite{Houde2019}: 
\begin{equation}
    E_{\mathcal{N}} = -\ln{\left[ e^{-2r}+(1-e^{-2r})\, \bar{\epsilon} + \tanh{(r)} \, \bar{\epsilon}^2 \, \delta^2\right]} \, ,
\end{equation}
where $\bar{\epsilon} = 1 - (\eta_s + \eta_i)/2$ indicates the average loss  and $\delta = (\eta_i - \eta_s)/(2 \bar{\epsilon})$ indicates the loss asymmetry \cite{Houde2019}.

% It should be noticed that also the purity of the state can be retrieved directly from the covariance matrix \cite{Adesso2005}:
%  \begin{equation}
%     \mu = \text{Tr}\rho^2 = \frac{1}{\sqrt{\text{Det}\mathbf{\sigma}}} \, .
% \end{equation}

%%%%%%%%%%%%%%%%%%%%%%%%%%%%%%%%%%%%%%%%%%%%%%%%%%%%%%%%%%%%%%%%%%%
\subsection{Collective quadratures and squeezing}
%%%%%%%%%%%%%%%%%%%%%%%%%%%%%%%%%%%%%%%%%%%%%%%%%%%%%%%%%%%%%%%%%%%
The collective quadrature operators can be defined as follows,
\begin{equation}
    \hat{x}_{+} = \left( \hat{x}_{s} + \hat{x}_{ i} \right) \, , \qquad \hat{x}_{-} = \left( \hat{x}_{s} - \hat{x}_{ i} \right)\, ,
\end{equation}
\begin{equation}
    \hat{p}_{+} = \left( \hat{p}_{s} + \hat{p}_{ i} \right) \, , \qquad \hat{p}_{-} = \left( \hat{p}_{s} - \hat{p}_{ i} \right)\, .
\end{equation}
For a two-mode squeezed state with squeezing parameter amplitude $r$ and accounting for loss with the beam splitter model as sketched in Fig \ref{setup_theory}, the variance of the collective quadrature is expected to be \cite{Houde2019}: 
% \begin{equation}
%     \braket{x_{+}^2} = \frac{1}{2} \, \left[ \left(1 -\frac{\eta_s +\eta_i}{2} \right) + \frac{e^{-2r}}{4} \left( \sqrt{\eta_s} +\sqrt{\eta_i}\right)^2 + \frac{e^{2r}}{4} \left( \sqrt{\eta_s} - \sqrt{\eta_i}\right)^2   \right]  \, .
% \end{equation}
% %%%%
\begin{equation}
\label{coll_var}
    \braket{x_{+}^2} = \frac{1}{2} \, \left[ \bar{\epsilon} + (1 - \bar{\epsilon} ) e^{-2r}   \right]  \, ,
\end{equation}
in the approximation of symmetric loss, $\eta_s = \eta_i = (1 - \bar{\epsilon})$. 

The amount of two-mode squeezing in dB is calculated as:
\begin{equation}
\label{coll_sq}
\text{Sq}_{+} = 10\log\left(\frac{\braket{x_{+}^2} }{0.5}\right) \, .
\end{equation}

For a fixed squeezing parameter amplitude $r$, corresponding to the gain value that experimentally maximizes the inferred logarithmic negativity ($G~=~1.30 \pm 0.05$), the amount of squeezing is limited by internal loss. 

\begin{figure*}[htb]
    \centering
    \includegraphics[width=0.5\textwidth]{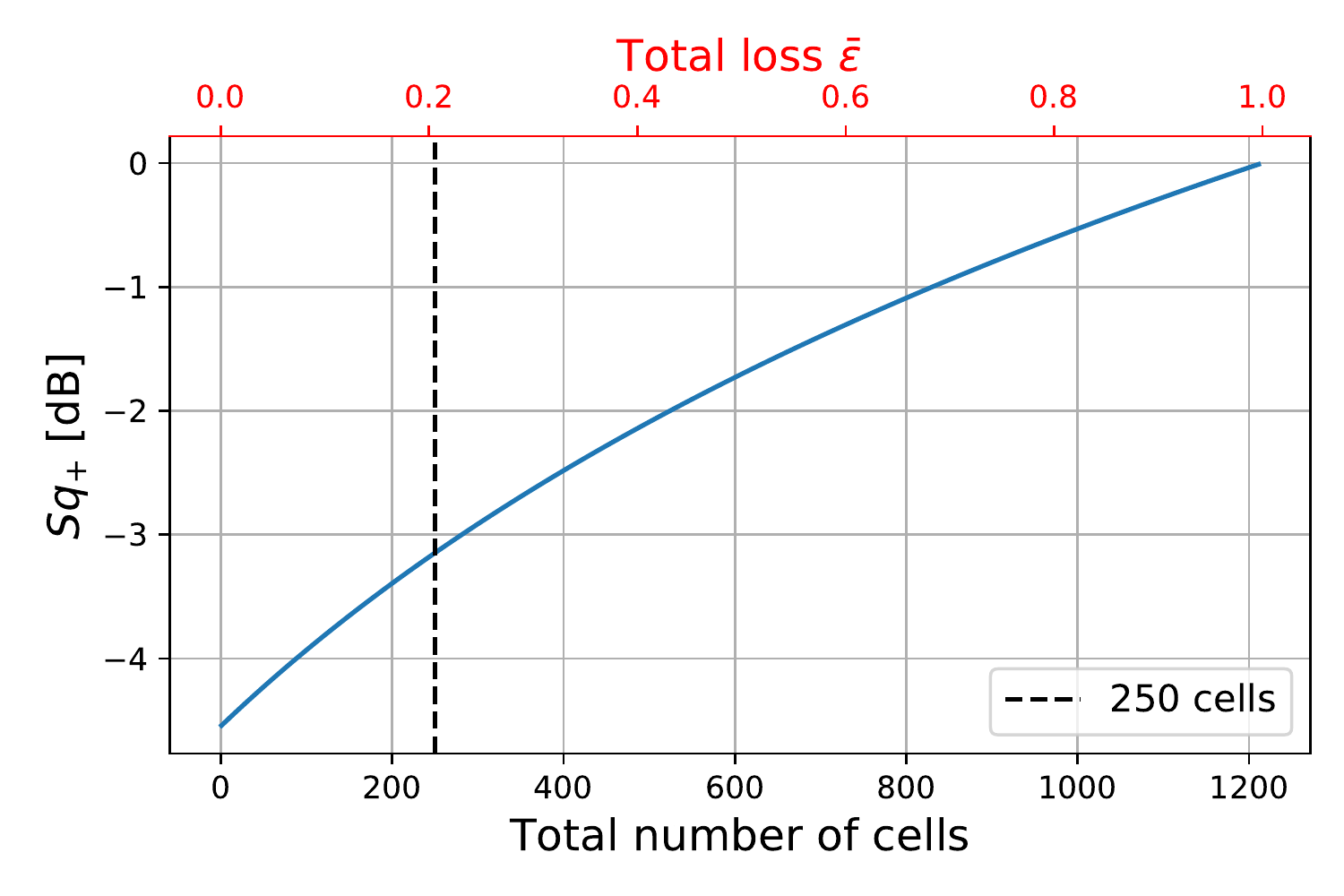}
    \caption{Amount of squeezing as a function of total number of unit cells (total loss $\bar{\epsilon}$). Fixed gain $G = 1.3$ ($r = 0.52$), fixed loss per unit length ($\bar{\epsilon}_{\text{cell}} = 8.2 \times 10^{-4}$) obtained from linear device characterization (section \ref{device}). The dashed line indicates our device length (250 cells).}
    \label{sq_vs_numb_cells_v1}
\end{figure*}

Knowing the loss per unit cell, % ($\bar{\epsilon}_{\text{cell}} = $),
given by the fabrication process (section \ref{device}), one simple way to mitigate the total loss $\bar{\epsilon}$ is to reduce the number of unit cells composing the device. In Fig \ref{sq_vs_numb_cells_v1}, we use equations \eqref{coll_var} and \eqref{coll_sq} to plot the predicted squeezing as a function of the total number of unit cells (total loss $\bar{\epsilon}$).
   %%%%%%%%%
This simple lumped element model for losses \cite{Houde2019} can be used to have a rough estimation of the amount of squeezing for a fixed TWPA gain value as a function of the device length (knowing the loss per unit cell). %In our case, we selected the number of unit cells to be 250 in order to aim for roughly -3 dB squeezing.}

%%%%%%%%%%%%%%%%%%%%%%%%%%%%%%%%%%%%%%%%%%%%%%%%%%%%%%%%%%%%%%%%%%%
\subsection{Rate of entangled bits}
%%%%%%%%%%%%%%%%%%%%%%%%%%%%%%%%%%%%%%%%%%%%%%%%%%%%%%%%%%%%%%%%%%%
% One of the key advantages of squeezing generation with TWPAs is the large bandwidth of such devices that can allow to reach large rates of entangled bits generation \cite{Flurin2012,Schneider2020}.\\
Defining the bandwidth $B$ as the frequency separation between the signal and idler frequencies, in our case $B = 2 \Delta$, the rate of entanglement bit generation is given by 
\begin{equation}
\text{Entanglement bit rate} = B \times E_{F},
\end{equation}
where $E_F$ is the entropy formation defined as
\begin{equation}
E_{F} = c_{+} \log_2{c_{+}} - c_{-} \log_2{c_{-}},
\end{equation}
with $c_{\pm} = (\delta^{-1/2} \pm \delta^{1/2})^2/4$ and $\delta = 2^{E_{\mathcal{N}}}$.\\
In our experiment, the limit to $B = 400$ MHz is given by the sampling rate of the acquisition board and not by the device itself. With a faster acquisition setup it could be possible to demonstrate even larger rate of entanglement bit generation.

%%%%%%%%%%%%%%%%%%%%%%%%%%%%%%%%%%%%%%%%%%%%%%%%%%%%%%%%%%%%%%%%%%%%%%%
\section{Pump power dependence}
\begin{figure*}[htb]
    \centering
    \includegraphics[width=0.5\textwidth]{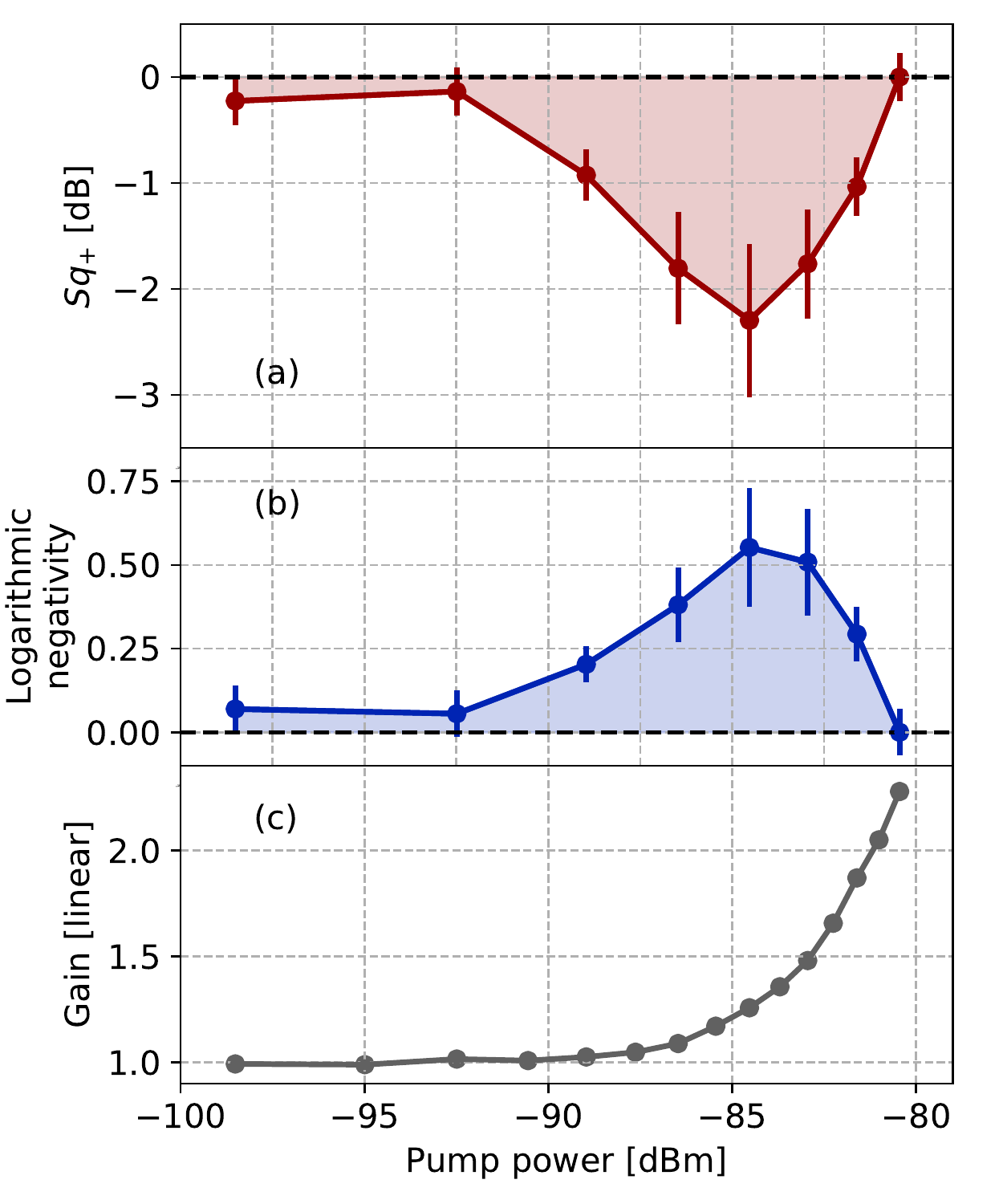}
    \caption{Squeezing (a), logarithmic negativity (b) and TWPA gain (c) as a function of the input pump power at the device. The inferred squeezing and logarithmic negativity are obtained for detuning $\Delta$ = 200~MHz, $N_{\text{rep}} = 10^7$ and pump frequency $f_p = 4.415$ GHz.}
    \label{sweep_pump_pow}
\end{figure*}

%%%%%%%%%%%%%%%%%%%%%%%%%%%%%%%%%%%%%%%%%%%%%%%
In the experiments shown in the main text, we optimize the pump power to get the maximum entanglement (maximum logarithmic negativity). This is reached for a linear gain $G~=~1.30 \pm 0.05$, corresponding to a squeezing parameter $r = \text{arcosh}(\sqrt{G}) = 0.50 \pm 0.04$. We experimentally observe that, for higher values of G, the inferred logarithmic negativity and squeezing are reduced. In Fig \ref{sweep_pump_pow} we show an example of experiment in which we measure TWPA gain, squeezing and logarithmic negativity as a function of the input pump power at the device.
\\
A detailed study of the effect of spurious non-linear processes in the high pump power (high gain) regime is beyond the scope of this work and will be subject of future investigations.

%%%%%%%%%%%%%%%%%%%%%%%%%%%%%%%%%%%%%%%%%%%%%%%%%%%%%%%%%%%%%%%%%%%%%%%
\section{Quadratures measurement and calibration}
%%%%%%%%%%%%%%%%%%%%%%%%%%%%%%%%%%%%%%%%%%%%%%%
A sketch of the signal and idler mode evolution from the generation to the detection is shown in Fig \ref{setup_theory}. 
%%%%%%%%%%
At the output of the TWPA, the signal and idler fields under investigation (described by the mode operators $\hat{a}_{s}$ and $\hat{a}_{i}$) go through a detection chain (yellow rectangle in the sketch) which starts at the output of the TWPA device and ends at the input of the acquisition board. The total system gain of such detection chain is indicated with $G_{\text{sys,} s}$ and $G_{\text{sys,} i}$ for signal and idler frequency respectively.\\
We can model the uncorrelated noise added by the amplification chain with a bosonic operator $\hat{h}_{s,i}$ associated with a thermal state. The final output field operators can be written as follows \cite{Lang2014}:
    \begin{equation}
        \hat{A}_{s} = \sqrt{G_{\text{sys,} s}} \,  \hat{a}_{s} + \sqrt{G_{\text{sys,} s}-1} \, \hat{h}_{s} \, ,  
    \end{equation}
    %%%%%%
    \begin{equation}
    \hat{A}_{i} = \sqrt{G_{\text{sys,} i}} \, \hat{a}_{i} + \sqrt{G_{\text{sys,} i}-1} \, \hat{h}_{i} ,
    \end{equation}
    %%%
where one can neglect the $-1$  in the square root, considering $G_{\text{sys,} s/i} \gg  1$.\\
What we experimentally measure are the four quadratures $\hat{X}_s$, $\hat{P}_s$, $\hat{X}_i$, $\hat{P}_i$, defined as:
    \begin{equation}
        \hat{X}_{s, i} = \frac{1}{2}\left( \hat{A}_{s, i} + \hat{A}^{\dagger}_{s, i} \right) \, , \qquad \hat{P}_{s, i} = \frac{1}{2 i}\left( \hat{A}_{s, i} - \hat{A}^{\dagger}_{s, i} \right).
    \end{equation}
%%%%%%%
In detail, we pump the TWPA with a pump frequency $f_p$ and measure the quadratures at signal frequency, $f_s = f_p + \Delta$, and idler frequency, $f_i = f_p - \Delta$. 
%%%%
The output voltage is analogically down converted to an intermediate frequency band (with $f_{\text{pump-IF}}$ being the central frequency) using a microwave mixer. The output from the mixer is then split in two channels and the two voltages are digitized every $0.5$ ns using two on board ADCs. The typical acquisition time is $\tau = 6 \mu s$. \\
The signals are then digitally down converted from the intermediate frequency $f_{{\text{pump-IF}}}-\Delta$ ($f_{{\text{pump-IF}}}+\Delta$) to DC, giving in output the raw quadratures $X_{s}^{\text{raw}}$, $P_{s}^{\text{raw}}$ ($X_{i}^{\text{raw}}$, $P_{i}^{\text{raw}}$) for the signal (idler), measured in  \textit{Volts}.\\
%%%%%
We repeat the quadrature acquisition $N_{\text{rep}}$ times, obtaining 6 two-dimensional quadrature histograms. We perform this measurement for both pump off and pump on.\\
We convert the measured raw quadratures from \textit{Volts} to dimensionless units by using the conversion factor 
$$
\gamma_{s, i} = \frac{1}{Z} \frac{\tau}{h (f_p\pm \Delta)}
$$ 
where $Z = 50 \,  \Omega$ and $h$ is the Plank constant. 
We get,  
$$
X_s = \sqrt{\gamma_s }\, X_{s}^{\text{raw}} \qquad ,  \qquad X_i = \sqrt{\gamma_i }\, X_{i}^{\text{raw}},
$$
$$
P_s = \sqrt{\gamma_s }\, P_{s}^{\text{raw}} \qquad , \qquad P_i = \sqrt{\gamma_i }\, P_{i}^{\text{raw}}.
$$
The final calibration step consists in considering the system gain (total gain of the detection chain). The final calibrated quadratures are given by:
$$
x_{s} = X_{s}/\sqrt{G_{\text{sys}, s}}  \qquad , \qquad p_{s} = P_{s}/\sqrt{G_{\text{sys}, s}}  \, ,
$$
%%%%%
$$
x_{i} = X_{i}/\sqrt{G_{\text{sys}, i}} \qquad , \qquad p_{i} = P_{i}/\sqrt{G_{\text{sys}, i}} \, ,
$$
where $G_{\text{sys}, s}$ and $G_{\text{sys}, i}$ are the system gain of the detection chain at signal and idler frequencies respectively. The system gain is obtained with a Shot Noise Tunnel Junction (SNTJ) calibrated noise source (see section \ref{SNTJ}).

%%%%%%%%%%%%%%%%%%%%%%%%%%%%%%%%%%%%%%%%%%%%%%%%%%%%%%%%%%%%%%%%%%%%%%%%
\section{Covariance matrix reconstruction}
%%%%%%%%%%%%%%%%%%%%%%%%%%%%%%%%%%%%%%%%%%%%%%%%%%%%%%%%%%%%%%%%%%%%%%%%%%

Assuming that the noise added by the amplification chain is thermal, i.e. it is associated with a diagonal covariance matrix $\mathbf{\sigma}_{\text{thermal}}$ \cite{Flurin2015}, the measured covariance matrices for pump on and off ($\mathbf{\sigma}^{\text{meas, on}}$, $\mathbf{\sigma}^{\text{meas, off}}$) can be then expressed as:
    \begin{equation}
    \mathbf{\sigma}^{\mathrm{meas, on}} = \mathbf{\sigma}_{\ket{\text{TMS}}}  + \mathbf{\sigma}_{\text{thermal}} +  \mathbb{1}_{4} \, , 
    \end{equation}
    \begin{equation}
    \mathbf{\sigma}^{\mathrm{meas, off}} =  \mathbf{\sigma}_{\ket{0}}  + \mathbf{\sigma}_{\text{thermal}} +  \mathbb{1}_{4} = \mathbf{\sigma}_{\text{thermal}} +  2 \, \mathbb{1}_{4} \, ,
    \end{equation}
where 
    \begin{equation}
    \mathbf{\sigma}_{\text{thermal}} = 
    \begin{pmatrix}
         2 \braket{\hat{h}^{\dagger}_{s} \hat{h}_{s}}_{\text{thermal}} & 0 & 0 & 0\\
        0 & 2 \braket{\hat{h}^{\dagger}_{s} \hat{h}_{s}}_{\text{thermal}} & 0 & 0\\
        0 & 0 & 2 \braket{\hat{h}^{\dagger}_{i} \hat{h}_{i}}_{\text{thermal}} & 0\\
        0 & 0 & 0 & 2 \braket{\hat{h}^{\dagger}_{i} \hat{h}_{i}}_{\text{thermal}} \, \\
        \end{pmatrix}. 
    \end{equation}
Finally, the covariance matrix of the two-mode quantum state generated at the output of the TWPA is given by \cite{Flurin2015}:
    \begin{equation}
        \mathbf{\sigma}_{\ket{\text{TMS}}} =  \mathbf{\sigma}^{\mathrm{meas, on}} -\mathbf{\sigma}^{\mathrm{meas, OFF}} +  \mathbb{1}_{4} \, .
    \end{equation}
                %

%%%%%%%%%%%%%%%%%%%%%%%%%%%%%%%%%%%%%%%%%%%%%%%%%%%%%%%%%%%%%%%%%%%%%%%%%%
\section{System gain calibration with SNTJ noise source \label{SNTJ}}
%%%%%%%%%%%%%%%%%%%%%%%%%%%%%%%%%%%%%%%%%%%%%%%%%%%%%%%%%%%%%%%%%%%%%%%%%%%

Here we describe the procedure used to estimate the system gain $G_{\text{sys}}$ using a calibrated noise source consisting in a shot noise tunnel junction (SNTJ) \cite{Spietz2003,Spietz2006,Chang2016}.  
The setup used for the calibration is shown in Fig \ref{setup}. 

We switch to the reference device (PCB), and measure the output noise power at a give frequency $f$ as a function of the bias voltage $V$ applied to the SNTJ.

We consider negligible loss for the PCB itself. In addition we use identical cables to connect the PCB and the TWPA to the switches.

The fraction of the noise power emitted by the SNTJ that is dissipated into a matched load in the quantum regime ($k_B T < hf$) is the following:
\begin{equation}
N =  \left[ \frac{1}{2} \left[\frac{e V + h f}{2 k_B} \coth{\left(\frac{eV + h f}{2 k_B T} \right)} + \frac{e V - h f}{2 k_B} \coth{\left(\frac{eV - h f}{2 k_B T} \right)}\right] + T_{\text{sys}} \right] \frac{1}{\tau} \, G_{\text{sys}} k_B \, ,
\label{SNTJ_formula}
\end{equation}
where the resolution bandwidth is defined as the inverse of the acquisition time $\tau$.

For a given frequency, we can fit the data using three free parameters: $T$, the SNTJ electronic temperature, the system noise temperature $T_{\text{sys}}$, and the system gain $G_{\text{sys}}$.  
This procedure allows the estimation of $G_{\text{sys}}$ at signal and idler frequencies. 

\begin{figure}[hbtp]
    \centering
    \includegraphics[width=0.7\textwidth]{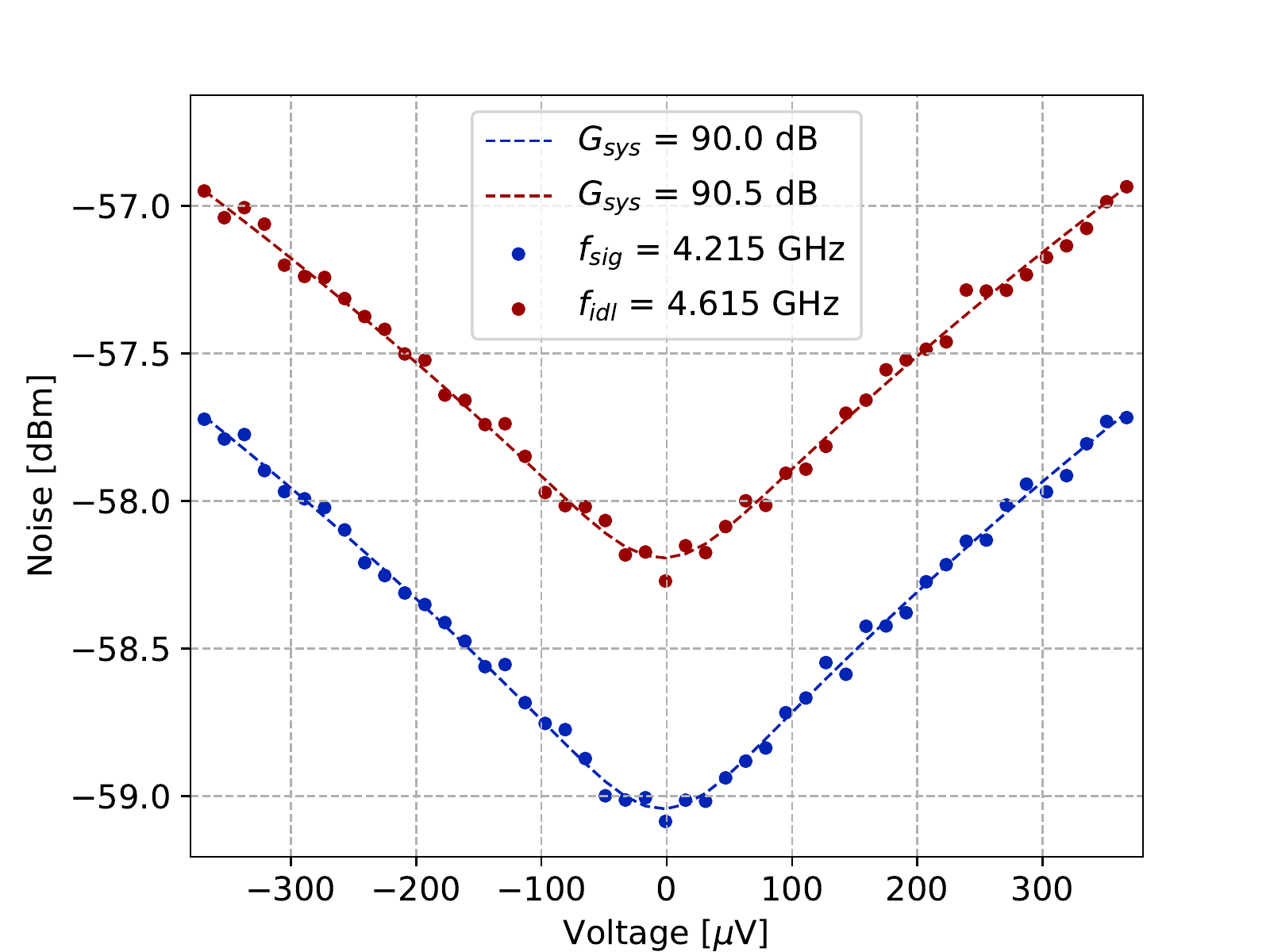}
    \caption{Measured output noise as a function of SNTJ DC voltage bias at signal and idler frequencies. Best fits are shown with dashed lines. The best fit parameters for $G_{\text{sys}}$ are reported in the legend.}
    \label{SNTJ_noise}
\end{figure}

In Fig \ref{SNTJ_noise} we show the SNTJ noise as a function of the bias voltage and the corresponding best fit for two given frequencies  corresponding to signal and idler for the data in Fig 2 of main text. 
This procedure is repeated for all the investigated signal and idler frequencies, allowing each time a careful estimation of  $G_{\text{sys}, s}$ and $G_{\text{sys}, i}$.%\\

It is important to notice that the values of $G_{\text{sys}, s/i}$, obtained from the best fit of the data with the expression in eq. \eqref{SNTJ_formula}, correspond to the total system gain, going from the reference plane of the SNTJ to the one of the room temperature acquisition board. It is necessary to correct such estimation for the loss occurring between the SNTJ and the input of the PCB. In particular, such loss are mainly due to SNTJ packaging, bias tee, directional coupler and cryo-switch insertion loss. In the frequency range of interest, the insertion loss due to SNTJ packaging including the bias tee has been reported to be 1 dB \cite{Chang2016}; in addition we consider the nominal maximum insertion loss of the used directional coupler and cryo-switch from the respective technical data sheet. In table \ref{tab:table1} we summarize the estimated maximum insertion loss contributions between SNTJ and PCB.
%%%%%%%%%%%%%%%%%%%%%%%%%%%%%%%%%%%%%%%%%%%%%
\begin{table}[h!]
        % \begin{center}
            \centering
            \caption{\label{tab:table1} Insertion loss budget bewteen SNTJ and PCB.}
            \begin{ruledtabular}
            \begin{tabular}{lcc} 
                     & Insertion Loss  &  Ref. \\ 
                \hline
                 SNTJ packaging + K250 bias tee & 1 dB & \cite{Chang2016} \\
                \hline 
                 directional coupler C20-0R518 & 0.7 dB & \cite{dir_coupler} \\
                \hline 
                 cryo-switch Radiall R591722600 & 0.3 dB & \cite{cryo-switch} \\
            \end{tabular}
            \end{ruledtabular}
        % \end{center}
    \end{table}
%%%%%%%%%%%%%%%%%%%%%%%%%%%%%%%%%%%%%%%%%%%%%

Thus, we estimate an upper bound of $2$ dB for the total insertion loss between the SNTJ reference plane and the device reference plane. The $G_{sys}$ value obtained from the SNTJ calibration is corrected by such upper bound loss estimation before it is used for the normalization of the quadrature data.

Considering that the bias tee model used in this work is different from the one in \cite{Chang2016} and considering typical ripples in the characterization of insertion loss of the used microwave components between SNTJ and device \cite{Malnou2021}, we assume an uncertanty of 1 dB in our final estimation of $G_{sys}$. The error bars on squeezing and logarithmic negativity in the main text and in Fig \ref{sweep_pump_pow} are obtained assuming that the total error is dominated by such uncertanty in the estimation of $G_{\text{sys}}$.

It should be noticed that, since we are using an upper bound estimation for the losses between SNTJ and device, we get an upper bound estimation for $G_{sys}$ and hence a lower bound estimation for the amount of entanglement and squeezing.

% \bibliography{C:/Users/martina.esposito/Dropbox/work/literature/bib_files/library.bib}% Produces the bibliography via BibTeX.
%\bibliography{zPaper_two_mode_squeezing.bib}% Produces the 
%